\documentstyle[12pt,aasms4,psfig]{article}

\def\ltsima{$\; \buildrel < \over \sim \;$}
\def\lsim{\lower.5ex\hbox{\ltsima}}
\def\gtsima{$\; \buildrel > \over \sim \;$}
\def\gsim{\lower.5ex\hbox{\gtsima}}

\begin{document}
\title
{SOURCE SIZE LIMITATION  \\ 
FROM VARIABILITIES OF A LENSED QUASAR}

\author{Atsunori Yonehara\altaffilmark{1,2}}

\altaffiltext{1}{Department of Astronomy, Kyoto University, Sakyo-ku, 
 Kyoto 606-8502, Japan}
\altaffiltext{2}{Research Fellow of the Japan Society 
 for the Promotion of Science}
\altaffiltext{3}{e-mail: yonehara@kusastro.kyoto-u.ac.jp}

\begin{abstract}

In the case of gravitationally-lensed quasars, 
it is well-known that there is a time delay between 
occurrence of the intrinsic variabilities in each split image.
Generally, the source of variabilities has a finite size, 
and there are time delays even in one image.  
If the origin of variabilities is widely distributed, 
say over $\gsim 100~{\rm pc}$ as whole, 
variabilities between split images will 
not show a good correlation even though their origin is identical. 
Using this fact, we are able to limit the whole source size of 
variabilities in a quasar below the limit of direct resolution 
by today's observational instruments. 

\end{abstract}


\section{INTRODUCTION}

Since Liebes (1964) and Refsdal (1964) have reported 
meaningful aspects of gravitational lensing phenomenon, 
many researchers rushed into the field of gravitational-lensing study, 
and presented many interesting results. 
This situation is not altered in these days. 

One of the most interesting gravitational-lens phenomena is quasar lensing.  
This is caused by a lensing galaxy (or galaxies) intervening 
observer and quasar. 
In the context of cosmology, it will be possible to estimate Hubble's constant 
from a time delay of the quasar variations  
between gravitationally-lensed, split images.  
The most successful study is by Kundi\'c et al. (1997, hereafter K97). 
They monitored Q0957+561 for a long time 
and performed robust determination of the time delay.
From their own result, they evaluate Hubble's ($H_{\rm 0}$) constant as 
$ 64^{+12}_{-13}~{\rm km~s^{-1}~Mpc^{-1} }$
based on the lens model constructed by Grogin and Narayan (1996, hereafter GN).

On the other hands, concerning the structure of quasar, 
we will discriminate the structure of central engine 
according to the effect of a finite source size.
Recently, Yonehara et al. (1998, 1999) performed 
realistic simulations of quasar microlensing, and 
showed that multi-wavelength observations will reveal the structure 
of accretion disk believed to be situated in the center of quasars.   
Furthermore, using precise astrometric technique, 
Lewis and Ibata (1998) indicated that it is also possible to probe 
the structure of quasar from image-centroid shift caused by microlensing.
Observationally, in the case of Q2237+0305, 
Mediavilla et al. (1998) detected a difference between
an extent of the continuum source and that of the emission-line source 
by two-dimensional spectroscopy, and  
limit the size of these regions.
Thus, quasar-lensing phenomena are a useful tool to probe 
not only for cosmology but also for the structure of quasar. 

Following these interesting researches, 
we propose a method to estimate, in this $letter$,  
the effect of a finite source size 
on time delays of the observed quasar variations  
between each gravitationally-lensed, split image, 
and to judge whether it is negligibly small or not 
and to limit the whole size of the source of quasar variability.
This is important because no such limitation has been done yet 
although the size of each variation, ``one shot'',   
had already been obtained order of days 
assuming causality in the individual source of variations.

In section 2, I describe the basic concept of this work,  
and simply estimate the time delay difference.
Next, I present some results of calculation 
for the case of Q0957+561 in section 3. 
Finally, section 4 is devoted to discussion.

\section{BASIC CONCEPT}

\begin{figure}[htbp]
\centerline{\psfig{figure=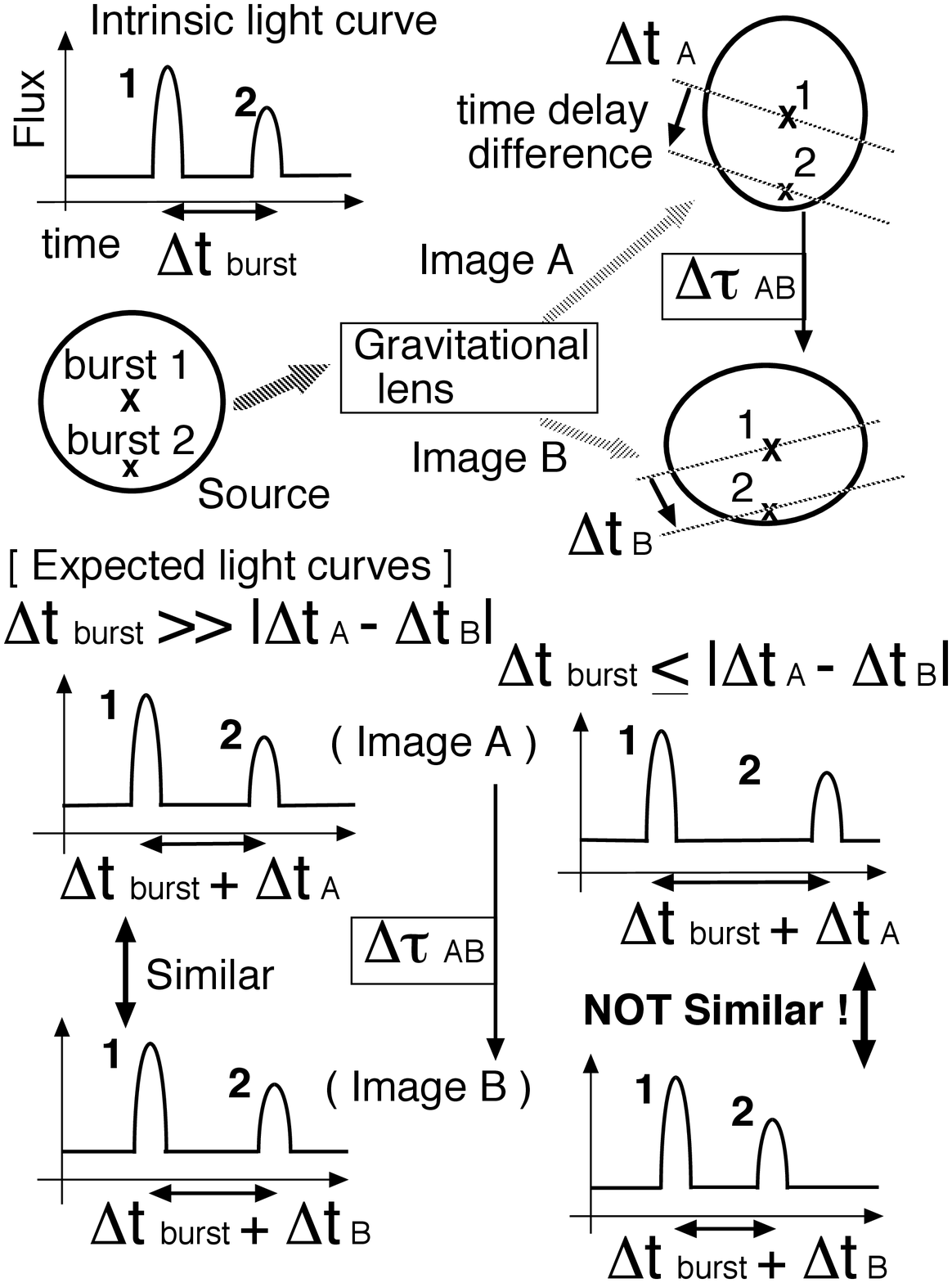,height=18cm}}
\caption{ 
Schematic view of the finite-source size effect for time delay 
between gravitationally-lensed, split quasar image (see section 2).
}
\label{tdsfig}
\end{figure}

The basic idea that we wish to present in this $letter$ is 
schematically illustrated in figure~\ref{tdsfig}. 
Suppose the situation that a quasar is macrolensed by lensing objects 
so that its image is split into two (or more) images.
The angular separation between these images 
is large enough to observe individually, 
say apparent angular separation is $\gsim 1~{\rm arcsec}$.
If we observe such quasar images, we will realize 
the intrinsic variabilities of quasar in each image 
as in the case of an ordinary, not gravitationally lensed quasar 
(e.g., recent optical monitoring results are shown in Sirola et al. 1998).
Because of the macrolensing effect, generally, the variabilities 
in such a quasar are not observed in both images at the same time. 
There is a time delay between these quasar images caused by 
a light path difference from the light path without lensing objects 
which originates from gravitational lens effect
(e.g., see Schneider, Ehlers, \& Falco 1992, hereafter SEF).
These facts are nicely demonstrated by K97. 

However, previous studies related to the time delay 
caused by gravitational lensing were not so much 
concerned with the source of variabilities, 
and the source of variabilities was treated as a point source. 
This treatment is reasonable, if the whole source size is 
negligibly small compared with the typical scale length 
over which a time delay changes. 
In contrast, actually, we only know that the source of quasar variabilities is 
smaller than the limit of the observational resolution, 
say $\lsim 1~{\rm arcsec}$ (e.g., in the case of {\it HST} observations, 
 $\sim 0.1~{\rm arcsec}$), and we do not know 
whether the whole source size is small or large 
compared with the scale length over which a time delay changes. 
Therefore, first, we should try to consider the effect of a finite source size 
on the expected observed light curve in quasar images. 

Then, if we include such an effect, what do we expect to see ?
The answer is easily understood by figure~\ref{tdsfig}. 
For simplicity, I consider only two images (image A and B) 
of the lensed quasar, 
and the source exhibit only two bursts 
(``burst 1'' and ``burst 2'', they occur in this order on the source plane)
with some time interval ($\Delta t_{\rm burst}$). 
The origin and separation of such bursts are not specified, 
we assume that these two bursts are not physically correlated, 
in other words they appear randomly. 
Additionally, we set a time delay difference 
between the position of the ``burst 1'' and the ``burst 2'' 
on image A as $\Delta t_{\rm A}$ and 
that on image B as $\Delta t_{\rm B}$.  

In the case of 
$\Delta t_{\rm burst} \gg |\Delta t_{\rm A} - \Delta t_{\rm B}|$, 
light curves of two images show apparently very similar feature,   
instead of its time delay at the very center.
Although the shape of light curves is altered from intrinsic one 
by the effect of finite source size 
as is depicted in lower left part of figure~\ref{tdsfig}, 
we can easily identify these two light curves are intrinsically the same one. 
Thus, we are able to obtain a robust time delay between two images.

On the other hands, in the case of 
$\Delta t_{\rm burst} \lsim |\Delta t_{\rm A} - \Delta t_{\rm B}|$, 
a previous fact does not hold any more. 
In this case, time interval between two bursts is significantly 
modified by the effect of its apparently large time delay difference 
($|\Delta t_{\rm A} - \Delta t_{\rm B}|$).
In such a situation, we can no longer conclude that 
light curves from two images have the same origin,  
even if we include an effect of time delay for the case of point source.  
We may seek for the reason for this to microlensing or something exotic. 
In other words, there will be no good correlation 
between light curves of two images. 
This is a serious problem not only to determine 
time delay or $H_{\rm 0}$ but also to construct a quasar structure, 
to determine the origin of variabilities, or some other problems.

Here, I will make a simple estimate of time delay difference between  
different parts of the source, i.e., the effect of a finite source.
In this estimate, I define ${\bf \beta}$,${\bf \theta_{\rm A}}$,
${\bf \theta_{\rm B}}$ as angular positions 
of the source center and those of the centers of two images. 
Therefore, ${\bf \theta_{\rm A}}$ and ${\bf \theta_{\rm B}}$ are 
the solutions of well-known lens equation (e.g., SEF), 
\begin{equation}
{\bf \beta} = {\bf \theta} - {\bf \alpha} ({\bf \theta}),
\label{eq:lens} 
\end{equation}
where, ${\bf \alpha}$ is a bending angle caused by intervening lens object(s), 
i.e., gravitational lens effect.
Furthermore, time delay from un-lensed light path ($\tau$) 
in the case of the image position is ${\bf \theta}$ and 
the source position is ${\bf \beta}$ is written as 
\begin{equation}
\tau = \frac{(1 + z_{\rm ol})}{2c} {\cal D} 
 \left| {\bf \theta} - {\bf \beta} \right|^2 
  - \frac{( 1 + z_{\rm ol})}{c^3} \Psi ({\bf \theta}).
\label{eq:delay}
\end{equation}
Here, $z_{\rm ol}$ is redshift from observer to lens, 
${\cal D}$ is effective lens distance that by using angular diameter distance 
from observer to lens ($D_{\rm ol}$), from observer to source ($D_{\rm os}$), 
and from lens to source ($D_{\rm ls}$),  
written as ${\cal D} = D_{\rm ol} D_{\rm os} / D_{\rm ls}$, 
and $\Psi$ is so-called ``effective lens potential'' (e.g., SEF).
Insert each image position into equation~(\ref{eq:delay}) 
and subtract one equation from the other, 
we obtain the well-known time delay expression between images A and B 
($\Delta \tau_{\rm AB}$),
\begin{equation}
\Delta \tau_{\rm AB}({\bf \beta}) = \frac{(1 + z_{\rm ol})}{2c} {\cal D} 
 \left( \left| {\bf \theta_{\rm A}} - {\bf \beta} \right|^2 -  
  \left| {\bf \theta_{\rm B}} - {\bf \beta} \right|^2 \right) - 
   \frac{( 1 + z_{\rm ol})}{c^3} \{ 
    \Psi({\bf \theta_{\rm A}}) - \Psi({\bf \theta_{\rm B}}) \} .
\label{eq:delayAB}
\end{equation}

Additionally, if we assume the position 
that is offset by $d{\bf \beta}$ from the center of the source 
and write $d{\bf \theta_{\rm A}}$ and $d{\bf \theta_{\rm B}}$ 
as image positions from the center of the image, these variables 
should fulfill the lens equation~(\ref{eq:lens}) again, i.e., 
\begin{equation}
({\bf \beta} + d{\bf \beta}) = ({\bf \theta}_i + d{\bf \theta}_i) 
 - {\bf \alpha} ({\bf \theta}_i + d{\bf \theta}_i),  
 ~ ~ ~ (i = {\rm A~or~B})
\label{eq:lensenh}
\end{equation}
or, subtracting this from equation~(\ref{eq:lens}) and 
adopt Taylor expansion to $\alpha$, 
we obtain another expression of equation~(\ref{eq:lensenh}), 
$ d{\bf \beta} = d{\bf \theta}_i - 
 \nabla_{\rm \theta}{\bf \alpha}({\bf \theta}_i)d{\bf \theta}_i + \cdots $.

Subtracting $\Delta \tau_{\rm AB}({\bf \beta})$ from 
$\Delta \tau_{\rm AB}({\bf \beta} + d{\bf \beta}) $, 
c.f., equation~(\ref{eq:delayAB}),
and using equation~(\ref{eq:lens}) and equation~(\ref{eq:lensenh}), 
we are able to obtain the time delay difference between 
the center of source and the other position offset by $d{\bf \beta}$
from the center of the source ($\delta \tau_{\rm AB} = 
\Delta \tau_{\rm AB}({\bf \beta} + d{\bf \beta}) - 
\Delta \tau_{\rm AB}({\bf \beta})$) on the source plane.

Moreover, by definition of effective lens potential, 
bending angle ${\bf \alpha}$ is related to the ${\bf \theta}$ through 
the derivative of effective lens potential $\Psi({\bf \theta})$ as 
$ \nabla_{\bf \theta} \Psi ({\bf \theta}) / {\cal D} c^2
 = {\bf \alpha}({\bf \theta})$.  
Since we are considering the origin of quasar variabilities, 
the source size is $\sim~{\rm kpc}$ at most and 
the distance from observer is typical cosmological scale $\sim 1~{\rm Gpc}$.
Thus, its apparent angular size is 
$\sim~{\rm kpc} / 1 {\rm Gpc} = 10^{-6}~{\rm rad}$. 
This seems to be small compared with image separation and  
the scale of bending angles which is typically a few arcsec. 
For $\delta \tau_{\rm AB}$, accordingly, 
we can adopt a Taylor expansion to ${\bf \alpha}$ and $\Psi$,  
neglect the higher terms than first order assuming 
$d{\bf \beta} \ll {\bf \beta}$ and $d{\bf \theta}_i \ll {\bf \theta}_i$.  
After using some algebra and putting $R$ as 
actual off-centered distance on the source plane, 
i.e., $R = |d{\bf \beta}| \cdot D_{\rm os}$, 
we are able to evaluate time delay difference as follows, 
\begin{eqnarray}
\delta \tau_{\rm AB} &\simeq& \frac{(1 + z_{\rm ol})}{2c}{\cal D} \{ 
  2{\bf \alpha}({\bf \theta_{\rm A}}) \nabla_{\bf \theta} {\bf \alpha}
   ({\bf \theta_{\rm A}}) d{\bf \theta_{\rm A}} - 
    2{\bf \alpha}({\bf \theta_{\rm B}}) \nabla_{\bf \theta} {\bf \alpha}
     ({\bf \theta_{\rm B}}) d{\bf \theta_{\rm B}} \} \nonumber \\
  &~& - \frac{(1 + z_{\rm ol})}{c^3} \{ 
   \nabla_{\bf \theta}\Psi({\bf \theta_{\rm A}})d{\bf \theta_{\rm A}} - 
    \nabla_{\bf \theta}\Psi({\bf \theta_{\rm B}})d{\bf \theta_{\rm B}} 
     \} \nonumber \\
 &=& \frac{(1 + z_{\rm ol})}{c}{\cal D} \left[ 
  \{ \nabla_{\rm \theta}{\bf \alpha}({\bf \theta_{\rm A}}) - 1 \}
   {\bf \alpha}({\bf \theta_{\rm A}})d{\bf \theta_{\rm A}} - 
    \{ \nabla_{\rm \theta}{\bf \alpha}({\bf \theta_{\rm B}}) - 1 \}
     {\bf \alpha}({\bf \theta_{\rm B}})d{\bf \theta_{\rm B}} \right] 
      \nonumber \\
 &\sim& \frac{(1 + z_{\rm ol})}{c}{\cal D} 
  \left| \left( {\bf \theta_{\rm B}} - {\bf \theta_{\rm A}} \right)
   \cdot d{\bf \beta} \right| \label{eq:direction} \\
 &\sim& 12 \left( \frac{1 + z_{\rm ol}}{2} \right) 
  \left( \frac{D_{\rm ol}}{D_{\rm ls}} \right) 
   \left( \frac{|{\bf \theta_{\rm B}} - {\bf \theta_{\rm A}}|}
    {1 {\arcsec}} \right) \left( \frac{R}{1~{\rm kpc}} \right) {\rm day}
\label{eq:delayest}
\end{eqnarray}
This one-dimensional evaluation is somewhat overestimated, however, 
for the calculations above, we did not use any restriction about 
lens model, and equation~(\ref{eq:delayest}) seems to be appropriate 
for any lens models and lensed systems 
except in some special situations, e.g., 
in the vicinity of caustics (or critical curves). 

Consequently, considering the fact that quasar optical intrinsic variabilities 
have timescale ${\rm day} \sim {\rm month}$, 
equation~(\ref{eq:delayest}) indicates that  
correlation between light curves of two images shown, in worst cases, 
will disappear, if the origin of quasar variabilities 
is extended over $\sim 1~{\rm kpc}$, 
i.e., maximum off-centered burst occurs at $\sim 1~{\rm kpc}$ from 
the center of quasar.

\section{EXAMPLES OF Q0957+561}

Finally, we will show some impressive result for the case of
Q0957+561 which is the first detected lensed quasar
by Walsh, Carswell, \& Weymann, (1979).

To demonstrate how the extended source effect works 
on the time delay determination in an actual lensed quasar, 
here, I will present simulation results of Q0957+561 as one example.
Using equation~(\ref{eq:delayest}), 
we are able to estimate a time delay difference 
between same source positions at different lensed images.
In this case, as is well known, if we use $z_{\rm ol} \simeq 0.36$, 
$z_{\rm os} \simeq 1.41$, $|\theta_{\rm B} - \theta_{\rm A}| \sim 6 {\arcsec}$ 
(e.g., GN) and assumed that 
$H_{\rm 0} \sim 60 {\rm km~s^{-1}~Mpc^{-1}}$, 
we will obtain $\delta \tau_{\rm AB} \sim 50~{\rm day}$
for the source with a size of $1~{\rm kpc}$ ! 

\begin{figure}[htbp]
\centerline{\psfig{figure=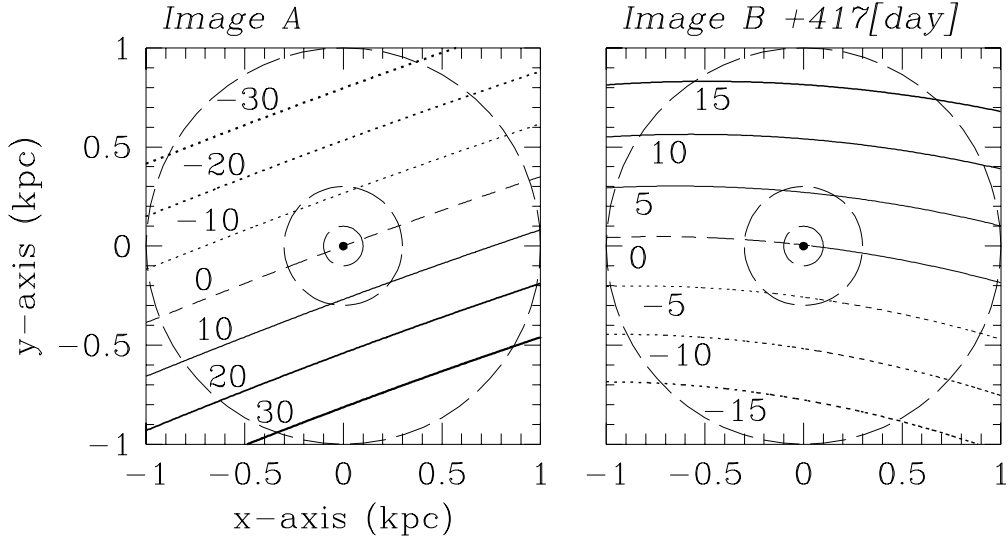,height=8cm}}
\caption{ Contour of time delay difference at image A (left panel) 
and image B (right panel) compared with their own center. 
The y-axis positive to north same as GN and 
the axis is almost parallel to 
the ${\bf \theta_{\rm A}} - {\bf \theta_{\rm B}}$ direction.
Source center is shown by filled circle at the center of each figures 
and long-dashed circles present a circle with a diameter of 
$1~{\rm kpc}$, $300~{\rm pc}$, and $100~{\rm pc}$ 
from outer one to inner one, respectively.
The time delay contours of leading parts (solid lines), 
preceding parts (dotted lines), and 
no time delay from the center parts (dashed lines) are presented. 
Concrete values are also shown in the figures. }
\label{tddiffig}
\end{figure}

Furthermore, to obtain more realistic results, 
we used isothermal SPLS galaxy with compact core 
as an example of lens model for Q0957+561 (details are shown in GN),  
adopted parameters listed in table 7 in GN as ``isothermal'' SPLS 
and calculated time delay difference between images center
and off-centered part of images ($\delta \tau_{\rm AB}$). 
For this calculation, we set $\Omega = 1.0$ for simplicity and 
took on convergence $\kappa \simeq 0.22$ and 
$H_{\rm 0} \simeq 64~{\rm km~s^{-1}~Mpc^{-1}}$ 
to reproduce the observed time delay following K97.
The resultant time delay contours compared with the image centers 
on the source plane are depicted in figure~\ref{tddiffig}.
On image A (left panel), a gradient of the contour is 
almost in the negative y-direction, although 
that of image B (right panel, this time delay advanced $\sim 417 {\rm day}$ ) 
is almost in the positive y-direction.
Additionally, time difference between the same position on the source 
reaches order of $\sim {\rm months}$ for 
the case of the source with a $\sim {\rm kpc}$ size, 
therefore, we expect disappearance of correlations 
between the light curves of image A and that of image B. 
Here, from equation~(\ref{eq:direction}), we can easily understand 
why do the contour lines show almost straight and 
perpendicular to y-direction. 
Product of $({\bf \theta_{\rm A}} - {\bf \theta_{\rm B}})$ and 
$d{\bf \beta}$ in equation~(\ref{eq:direction}) means that 
time delay difference determined mostly by 
the element which parallel to $({\bf \theta_{\rm A}} - {\bf \theta_{\rm B}})$ 
of displacement $d{\bf \beta}$. 
Therefore, the time delay difference significantly alters 
along the $({\bf \theta_{\rm A}} - {\bf \theta_{\rm B}})$ direction 
and almost constant along the perpendicular direction.  

\begin{figure}[htbp]
\centerline{\psfig{figure=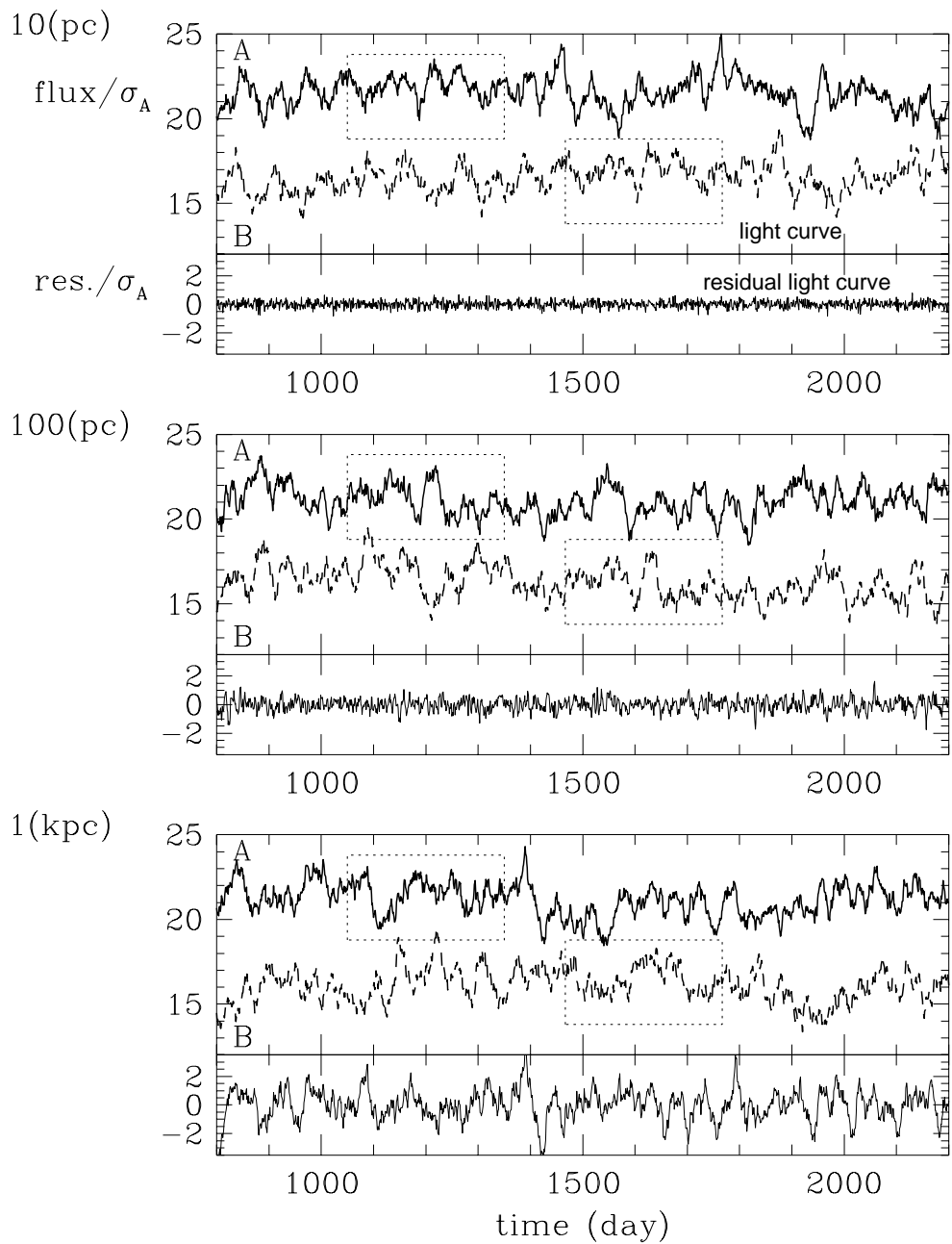,height=18cm}}
\caption{Simulated light curves (upper one of each panel) 
of image A (solid curve) and B (dashed curve),  
and its residual light curves (lower one) for some different source size. 
Numeric values on the horizontal line are fluxes normalized 
by a standard deviation of light curve of image A ($\sigma _{\rm A}$).  
The whole source sizes are $10~{\rm pc}$ (upper panel), 
$100~{\rm pc}$ (middle panel), and $1~{\rm kpc}$ (lower panel).
The curves are arbitrary shifted in the vertical direction. 
A light curve in dotted box of image A and B 
should be the same in the case of point-source treatment.}
\label{tdlcfig}
\end{figure}

Moreover, we simulated expected light curves of variabilities 
in both quasar images using superposition of simple bursts 
with triangled shape and duration of $\sim 10$ days 
which are randomly distributed in time, in space, and in amplitude.  
For the whole source size, I consider three cases, 
$1~{\rm kpc}$, $100~{\rm pc}$ and $10~{\rm pc}$.
Using the same procedure to produce figure~\ref{tddiffig}, 
we calculated time delay from the center of image A over both images, 
randomly produce bursts, sum up all bursts 
and finally obtain expected light curves as presented in figure~\ref{tdlcfig}. 
``Residual light curves'' produced by subtracting properly-shifted 
light curve of image B from that of image A, are also shown in the figure.
In the case of the smallest source, $R = 10~{\rm pc}$, 
and still in the case of middle source size, $R = 100~{\rm pc}$, 
we can easily recognize the coherent pattern 
in the light curves of images A and B but with time delay of  
$\sim 417~{\rm days}$ advanced light curve of image B. 
Time delay between two image centers is able to be determined fairly well. 
However, in the case of largest source, $R = 1~{\rm kpc}$, 
it seems no correlation between two light curves 
even if we already know the time delay between two image centers, 
and we may misunderstand that the variabilities 
did not originate in source itself !
This feature is far from the observed properties 
that the time delay between two images is determined easily 
even if we fit them by eyes. 
Therefore, I conclude that the size of source that is origin of 
quasar variabilities should be smaller than $100~{\rm pc}$, 
namely, maximum acceptable size is order of $\sim 10~{\rm pc}$ 
from this simple simulation.

\section{DISCUSSIONS AND COMMENTS}

As we examined, if we include the finite source-size effect to 
the time delay determination from quasar variabilities, correlation between 
expected light curves of each lensed image will disappear 
in the case of the size is sufficiently large, say $\sim 1 {\rm kpc}$.
Using this fact, we can limit the size of the region 
where quasar variabilities are produced, 
from the correlation between light curves of 
multiple lensed quasar images. 
Furthermore, since the size of the origin of intrinsic quasar variability 
reflects a physical origin of the variabilities,  
we can also determine the origin of the variabilities,  
e.g., whether it is disk instability (Kawaguchi et al. 1998) or 
star burst (Aretxaga, Cid Fernandes, \& Terlevich, 1996).
Particularly, in the case of Q0957+561, 
the origin of variabilities has a size smaller than $100~{\rm pc}$. 
This value is consistent with disk instability model, 
because of its small size ($\sim 0.01 {\rm pc}$ for 
$1000$ Schwarzschild radius accretion disk
surrounding $10^{8}M_{\odot}$ supermassive black hole).
Starburst model can be rejected, 
since starburst region is $100~{\rm pc} \sim 1~{\rm kpc}$.
Hence, for the origin of intrinsic quasar variabilities,
the disk instability model is more preferable, as was   
indicated by Kawaguchi et al. (1998) already.
To draw this conclusion more critically, 
we should do this study more precisely in future.

Additionally, the fact that a larger source size tends to reduce 
a good correlation between the light curves of each image provides 
an answer to the question why time delay determination 
from radio flux gave a wrong answer except recent works, 
e.g., Haarsma et al (1999). 
Generally, radio emitting region is believed to   
have a larger size than that of optical photon  
because of the existence of large radio lobe and/or jet component, 
and the effect we have shown in this $letter$ may be significant.  
Thus, robust determination of the time delay seems to be difficult.  

If such a effect is significant in the well-known lensed quasar Q2237+0305, 
microlens interpretation of individual variabilities  
(e.g., see Irwin et al. 1989) will be rejected. 
Fortunately, however, this may not be the case 
because for this source, caused by its quite nice symmetry of lensed image, 
the effect seems not to be so significant and 
intrinsic variabilities will be expected to appear 
in every images with good correlations.

If we develop this technique furthermore, 
and adapted to another multiply-imaged lensed quasar, 
we will determine the size of the most interesting part in quasar.

\acknowledgements

The author would like to express his thanks to 
Toshihiro Kawaguchi, Jun Fukue 
for their number of comments,  
and to referee for his/her meaningful suggestions.
The author also grateful to Shin Mineshige for his valuable discussions 
and kind reading of the previous draft.
This work was supported in part by the Japan Society for
the Promotion of Science (9852).

\clearpage

\end{document}